\documentclass[twocolumn,prl,showpacs]{revtex4}

\usepackage{epsfig}

\usepackage{amsmath}
\usepackage{amsfonts}
\usepackage{amssymb}

\begin{document}

\title{Spatial small-world networks: a wiring-cost perspective }
\author{Thomas Petermann}
\email{Thomas.Petermann@alumni.ethz.ch}
\author{Paolo De Los Rios}
\address{Institute of Theoretical Physics, LBS, Ecole Polytechnique F\'ed\'erale de Lausanne - EPFL, CH-1015 Lausanne, Switzerland.}
\date{\today}

\begin{abstract}

Supplementing a lattice with long-range connections effectively models small-world networks characterized by a high local and global interconnectedness observed in systems ranging from society to the brain. If the links have a wiring cost associated to their length $l$, the corresponding distribution $q(l)$ plays a crucial role. Uniform length distributions have received most attention despite indications that $q(l) \sim l^{-\alpha}$ exist, e.g.\ for integrated circuits, the Internet and cortical networks. Here we discuss for such systems the emergence of small-world topology, its relationship to the wiring costs, the distribution of flows as well as the robustness with respect to random failures and overload. The main finding is that the choice of such a distribution leads to favorable attributes in most of the investigated properties.

\end{abstract}
\maketitle

The explosion of research activity in the field of complex networks has led to a novel framework in order to 
describe systems in disciplines ranging from the social sciences to biology \cite{cnet}. 
One feature shared by most real networks is the {\it small-world} (SW) property, namely, every pair of nodes is 
separated, on the average, by only a few links \cite{watts}. Although the SW phenomenon has first been introduced 
in a social context \cite{milgram}, it is also relevant for
communication and technological systems such as the Internet \cite{pastoryook} or electronic circuits \cite{cancho}. 
A more precise definition of the SW topology implies not only a short distance between every pair of nodes,
but {\it also} a high degree of local interconnectedness, that is, for every node, most nodes close to it should also 
be close to each other.
Such properties are of great relevance for communication systems:
SW networks are particularly efficient for message passing protocols that 
rely only on the local knowledge of the network available to each node \cite{kleinberg01}.
It has also been recently pointed out that SW networks could describe the architecture of neural networks: 
{\it in vitro} neuronal networks \cite{shefi}, brain functional networks \cite{eguiluz} as well as the cerebral 
cortex \cite{sporns} exhibit SW features. In fact,the topology plays a crucial role in a neural network, since 
the high local interconnectedness gives rise to coherent oscillations while short global distances ensure a fast system 
response \cite{lago}. 

Network possessing both of the relevant SW properties can be realized 
as follows starting from a regular (or homogeneous) distribution
of nodes in space, with only nearest neighbor connections (the nearest 
neighbors are simply defined on a lattice, and can be defined via Voronoi
tessellations for random node distributions \cite{voronoi}): 
every node can establish one long-range connection ({\it shortcut})
over the network
with probability $p$. In this model, $p$ allows to interpolate 
continuously between a fully regular ($p=0$) and an entirely random ($p=1$) 
topology, the precise nature of this transition being discussed below. 
This "re-wiring" procedure can be accompanied
by a contemporary loss of one local connection of the re-wired node
\cite{watts}, without significant changes in the way the small-world topology emerges.
We therefore deal with the model where re-wiring is not accompanied by edge removal.

In the original formulation of the SW model, 
which received most of the attention \cite{swrev}, 
the length distribution of the shortcuts is uniform, since a node can choose any
other node for re-wiring, irrespective of their Euclidean distance. Yet, new interesting properties emerge if this condition is relaxed, for example if the distribution $q(l)$ of connection lengths $l$ decays as a power law, $q(l)\sim l^{-\alpha}$. 
The navigability in 
such a small-world, for example, depends on the corresponding decay exponent 
\cite{kleinberg}, and the nature of random walks over the network is also affected 
\cite{blumen}. It was even conjectured that the fundamental mechanism behind the SW phenomenon is neither disorder nor randomness, but rather the presence of multiple length scales \cite{kasturirangan} in agreement with $q(l) \sim l^{-\alpha}$.

Moreover, real-world networks are unlikely modeled {\it \`a la} Watts and Strogatz:
if shortcuts have to be physically
realized, the cost of a long-range connection is likely to grow with its length. Since 
nodes connected by shortcuts can be at any Euclidean distance from each other, it turns out
that the amount of resources that they have to invest in their connections
grows linearly with the system linear size, and it is, {\it a priori}, impredictable.
This is far from optimal for systems composed by entities with limited resources
({\it e.g.}, providers or neurons).
Indeed, local (single node) and global {\it wiring cost} considerations are likely to be
key factors in the formation of real SW networks \cite{chklovskii,laughlin,klyacho,buzsaki,sporns2, yook}. Regarding connection-length distributions $q(l) \sim l^{-\alpha}$, such measurements were reported for systems created through self-organization, design and evolution, namely for the Internet \cite{yook}, integrated circuits \cite{payman}, the human cortex \cite{schuez} and for regions of the human brain correlated at the functional level \cite{eguiluz}. Some modeling effort taking into account the constraint of wiring minimization has been made for systems where the connection lengths are \cite{karbowsky} or are not distributed according to a power law \cite{barth,kaiser,gastner}, and such length distributions emerge quite naturally when wiring costs along with shortest paths are minimized \cite{mathias}.

In this Letter we re-analyze the SW phenomenon from a wiring cost perspective, for networks in $D$ dimensions,
built
using a power-law decaying distribution of shortcut lengths. We find, both analytically and numerically,
that $\alpha<D+1$ is the condition for the emergence of SW behavior.
We also show that, given a fixed total wiring cost, networks with larger values of $\alpha$ are both
smaller worlds and more robust upon random removal of shortcuts.

We construct our SW networks as follows. Given a $D$-dimensinal lattice ($D=1,2$) of linear size $L$, consisting of $N=L^D$ sites, subject to periodic boundary conditions, we equip it with $pN$ additional connections. For every such link, we first choose its length according to the distribution $q(l) \sim l^{-\alpha}$ and then put it on the lattice by randomly choosing one endpoint and the other at distance $l$, such that no pair of sites is connected by more than one additional connection.
For small values of $p$, our formulation of the model corresponds to the case where at every site, a link is added with probability $p$ - the other endpoint being chosen as above. Yet, it has the advantage that $p$ can be greater than 1 which allows us to look beyond the simple probabilistic version.
\begin{figure}[b]
\includegraphics[width=8cm]{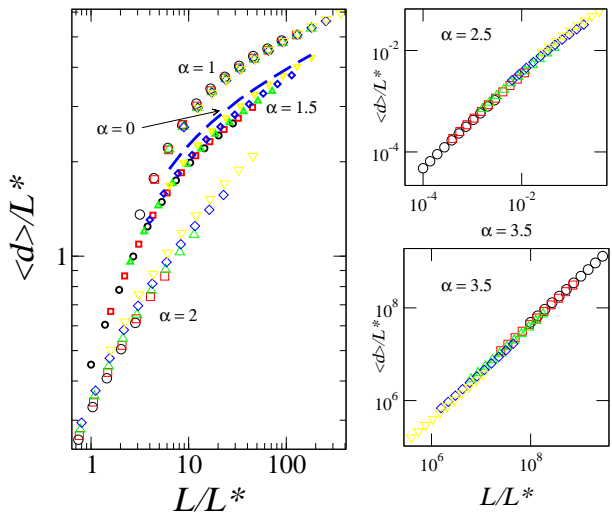}
\caption{Mean distance versus linear system size, both of these quantities being rescaled by $L^*_\alpha(p)$, for $p=0.001$ ($\circ$), $p=0.002$ ($\square$), $p=0.004$ ($\triangle$), $p=0.008$ ($\diamond$) and $p=0.016$ ($\triangledown$). The exponent $\alpha$ ranges from 0 to 3.5 as indicated. The data collapses confirm Eqs.\ (\ref{equ:SW_cond}) and (\ref{equ:scal}) also for $\alpha>\alpha_c$ (lower right panel).}
\label{fig:2D_coll}
\end{figure}
Clearly, a certain amount of real shortcuts, i.e.\ long additional links, is required for SW topology to emerge. We therefore expect that the length-distribution exponent $\alpha$ has to be smaller than a critical value $\alpha_c$. In order to find this condition, we look at the probability that an arbitrarily chosen additional link is a real shortcut
\begin{equation}	\label{equ:int_c}
P_c(L)=\int_{(1-c)L/2}^{L/2} q(l) dl,
\end{equation}
c being small but finite, and require the presence of the order of one such connection:
\begin{equation}	\label{equ:arg}
P_c(L) p^*(L) L^D \simeq 1
\end{equation}
where $p^*(L) L^D$ is the desired number of shortcuts, implying the emergence of SW topology for $p \gg p^*(L)$. After evaluating the scaling of the integral (\ref{equ:int_c}), Eq.\ (\ref{equ:arg}) reads
\begin{equation}	\label{equ:SW_cond}
p^*(L) \sim \begin{cases}
L^{-D} \qquad\qquad &\text{if} \quad \alpha<1,\\
\ln(L)/L^D \qquad &\text{if} \quad \alpha=1,\\
L^{\alpha-D-1} \qquad\quad &\text{if} \quad \alpha>1.
\end{cases}
\end{equation}
Eq.\ (\ref{equ:SW_cond}) implies the presence of a unique length scale $L^*_\alpha(p)$, suggesting that the mean distance can be expressed as
\begin{equation}	\label{equ:scal}
\langle d \rangle = L^* \mathcal F_\alpha \Bigl(\frac{L}{L^*}\Bigr),
\end{equation}
the scaling function obeying \cite{barth2,newman1}
\begin{equation}	\label{equ:scal_func}
\mathcal F_\alpha (x) \sim \begin{cases}
x \qquad\qquad &\text{if} \qquad x \ll 1,\\
\ln(x) \qquad &\text{if} \qquad x \gg 1 \end{cases}
\end{equation}
where the second line of Eq.\ (\ref{equ:scal_func}) may also read $[\ln(x)]^{s(\alpha)}, s(\alpha)>0$.
In other words, SW topology implies that the mean distance increases only logarithmically with the system size ($L \gg L^*$) whereas in a {\it large world} (LW), $\langle d \rangle \sim L$, i.e.\ if $L \ll L^*$. For $\alpha<1$, Eq.\ (\ref{equ:SW_cond}) yields $L^*(p) \sim p^{-1/D}$, a result which was previously derived for the special case $\alpha=0$ \cite{newman1,demenezes}. The characteristic length scales as $L^*(p) \sim p^{1/(\alpha-D-1)}$ for $\alpha>1$, thus becoming infinite at $\alpha_c=D+1$. We therefore have two possible regimes for $\alpha<\alpha_c$ while LW behavior prevails for $\alpha \ge \alpha_c$. Fig.\ \ref{fig:2D_coll} shows the rescaled mean distances as a function of the rescaled linear system size for diffent values of $\alpha$ and $p = 0.001, 0.002, ...,0.016$ in each set for the case $D=2$. The observed data collapses for all the length-distribution exponents confirm Eq.\ (\ref{equ:SW_cond}) obtained by our simple argument as well as Eq.\ (\ref{equ:scal}). We numerically verified Eq.\ (\ref{equ:scal_func}), especially in the limit $L/L^* \ll 1$, the logarithmic tail of $\mathcal F_\alpha$ further being exhibited best for small $\alpha$. Previously, $\alpha_c=D+1$ was derived by defining SW topology via the scaling of the clustering coefficient rather than by means of the scaling of distances \cite{sen}. Our argument can be generalized for a model where a shortcut between any pair of sites ${\bf x}$ and ${\bf y}$ is added with a probability proportional to $|{\bf x}-{\bf y}|^{-\alpha}$, finding $\alpha_c=2D$ in agreement with \cite{biskup}, which can also be derived through a rescaling argument \cite{moukarzel}.

\begin{table}[t]
\caption{Behavior of the moments of the shortcut-length distribution as a function of the linear system size $L$.}
\begin{ruledtabular}
\begin{tabular}{ccccc}
 & $0 \leq \alpha<1$ & $1<\alpha<2$ & $2<\alpha<3$ & $\alpha>3$  \\
\hline
$\langle l \rangle$ & $L$ & $L^{2-\alpha}$ & const & const \\
$\langle l^2 \rangle$ & $L^2$  & $L^{3-\alpha}$  & $L^{3-\alpha}$  & const
\end{tabular}
\end{ruledtabular}
\label{tab:moments}
\end{table}
The moments $\langle l \rangle$ and $\langle l^2 \rangle$ play a crucial role as far as the wiring costs of the networks are concerned. The total wiring cost $C_W =p L^D \langle l \rangle$ is also an important quantity, its minimisation governing, for example, the evolution of cortical networks \cite{laughlin}. We find for the first two moments the scaling relations summarised in Tab.\ \ref{tab:moments}, the expressions for integer $\alpha$ being modified by logarithmic corrections. In 2 dimensions, SW topology can be realized even if $\langle l \rangle = \text{const}$ (that is, for $2< \alpha < 3 = \alpha_c$) whereas this is not the case in 1 dimension where $\langle l \rangle$ becomes finite in the $L \rightarrow \infty$ limit only above $\alpha_c=2$. Moreover, if $D=3$, it is even possible to have $\langle l \rangle = \mathcal O (1) = \langle l^2 \rangle$ while still being in the SW regime for $3 < \alpha< 4 = \alpha_c$. An appropriate choice of the parameters $D$ and $\alpha$ is thus the key to constructing networks which are both efficient (SW topology) and economical (low wiring costs). 

\begin{figure}[b]
\includegraphics[width=8.5cm]{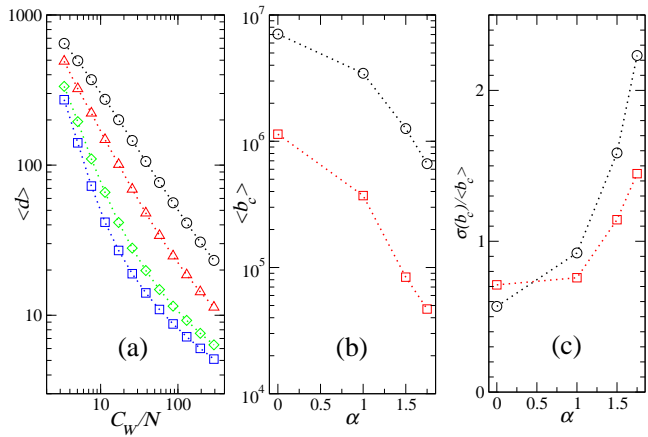}
\caption{(a) Mean distance as a function of the total wiring costs (divided by the number of sites) for 1 dimensional topologies. The curves ($\circ$: $\alpha=0$, $\triangle$: $\alpha=1$, $\diamond$: $\alpha=1.5$ and $\square$: $\alpha=1.75$) show that the mean distance decreases with $\alpha$ for a fixed value of $C_W/N$. (b) Mean values and (c) relative standard deviations of the link betweenness centrality at wiring costs $C_W/N=10$ (circles) and $C_W/N=100$ (squares). The results shown in these 3 panels were obtained for lattices consisting of $10^4$ sites and by averaging over 100 realizations of the randomness.}
\label{fig:wir_costs}
\end{figure} 
It is furthermore interesting to have a closer look at the relationship between the wiring costs and the topology. As $\alpha$ varies, one can ask what mean distance results given a total amount of wiring length for the additional connections (i.e.\ the total cost). Fig.\ \ref{fig:wir_costs}a reports these dependencies for $\alpha=0,1,1.5$ and 1.75 (going from the uppermost to the lowest set) for 1 dimensional topologies of $10^4$ sites. The largest value of $\langle d \rangle$ (the leftmost circle) corresponds to the length scale $L^*<10^3 \ll 10^4=L$, thus all the points in the figure represent the system in the SW regime. It can clearly be seen that the mean distance decreases with $\alpha$ at fixed wiring costs $C_W/N$, i.e.\ the larger $\alpha$ the smaller the world. This behavior is qualitatively recovered when expressing Eq.\ (\ref{equ:scal}) in terms of $x=C_W/N=p\langle l \rangle$. We made similar observations in 2 dimensions.

In order to obtain a more complete picture, we also studied the implications of our model regarding the distribution of flows over the additional links. The corresponding quantity is the link betweenness centrality which is defined as 
\begin{equation} \label{equ:bc}
b_c(s)=\sum_{A,B} \frac{n_{AB}(s)}{n_{AB}}
\end{equation}
where $n_{AB}$ is the number of shortest paths between nodes $A$ and $B$, and $n_{AB}(s)$ counts only those going through connection $s$, the sum running over all pairs of nodes \cite{freeman}. By using an efficient algorithm for the $b_c$-computation \cite{brandes}, we obtained for 1 dimensional topologies of $10^4$ nodes with varying $\alpha$ but again fixed wiring costs the mean values $\langle b_c \rangle$ and relative standard deviations $\sigma(b_c)/\langle b_c \rangle$ (inset) shown in Figs.\ \ref{fig:wir_costs}b and \ref{fig:wir_costs}c (upper sets: $C_W/N=10$, lower sets: $C_W/N=100$). As Eq.\ (\ref{equ:bc}) can be transformed into $\langle b_c \rangle \sim \langle d \rangle / E$ \cite{dimitrov}, $E$ being the total number of links, and if the role of the links of the underlying lattice is ignored, $\langle b_c \rangle$ clearly decreases with $\alpha$ since both $\langle d \rangle$ and $1/E \sim (pN)^{-1}=\langle l \rangle/C_W$ decline with $\alpha$ at constant costs in agreement with the numerical findings. The fact that the higher $\sigma(b_c)/\langle b_c \rangle$ the larger $\alpha$ indicates that the extent to which shortest paths overlap is an increasing function of $\alpha$ as $\langle b_c^2 \rangle$ relates directly - via Eq.\ (\ref{equ:bc}) - to this topological property. This is very reasonable since the higher $\alpha$ the smaller the fraction of real shortcuts, i.e.\ those carrying the traffic.

\begin{figure}[t]
\includegraphics[width=8cm]{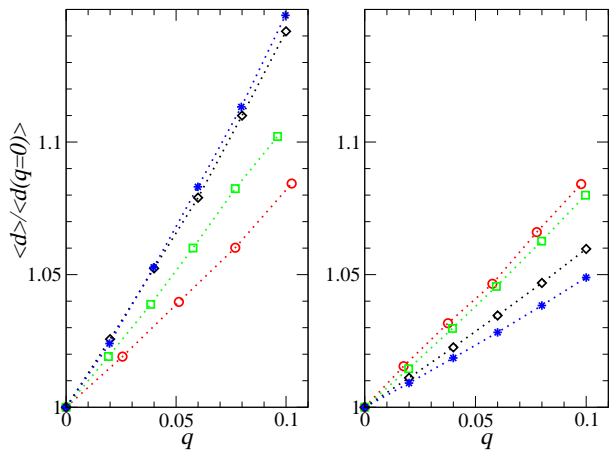}
\caption{Robustness with respect to random failures. These figures show that the mean distance increases when a fraction $q$ of the shortcuts is removed. The initial 1D networks have a fixed cost: $C_W/N=10$ (left) and $C_W/N=100$ (right), and $\alpha$ takes on the values 0 ($\circ$), 1 ($\square$), 1.5 ($\diamond$) and 1.75 ($\star$). These results were obtained by averageing over 1000 realizations of networks consisting of $L=10^4$ nodes.}
\label{fig:rf}
\end{figure}
There is another important property, namely the behavior of the networks with respect to link deletion. The associated concept, that is, robustness, can for example be defined as the extent to which the mean distance increases when a definite fraction of additional links is failing. Two different types of failure can be distinguished. On the one hand, it is possible that certain connections are malfunctioning for whatever reason, resulting in {\it random failures}. The other scenario is directly related to the traffic on the network: in the most simple situation, every node sends one packet to all other nodes implying different intensities of flows through the various additional connections. As the latter may only be able to carry flow amounts which lie below a certain threshold, {\it overload} of the long-range links occurs which is the second type we shall investigate.

\begin{table}[b]
\caption{Overload-related robustness (simultaneous removal of $10\%$ of the most charged shortcuts). The values shown here were obtained by simulating the process for 100 different realizations of networks consisting of $N=100 \times 100$ sites with initial wiring costs $C_W/N=10$. The emerging picture is that networks with low $\alpha$ are most robust.}
\begin{ruledtabular}
\begin{tabular}{ccccc}
$\alpha$ & 0 & 1 & 2 & 3 \\
\hline
$\langle d \rangle/\langle d(q=0)\rangle$ & $1.05 \pm 0.01$ & $1.08 \pm 0.01$ & $1.24 \pm 0.02$ & $1.37 \pm 0.04$ \\
\end{tabular}
\end{ruledtabular}
\label{tab:ol}
\end{table}
As far as random failures are concerned, Fig.\ \ref{fig:rf} illustrates to what extent the mean distances of $1D$ networks increase for $C_W/N=10$ (left) and $C_W/N=100$ (right), $\alpha$ taking the values $0,1,1.5$ and 1.75 if up to $10\%$ of the additional links are deleted. We find that for low costs ($C_W/N=10$), $\alpha=0$ corresponds to the most robust system, becoming more fragile as $\alpha$ increases. Yet, at high wiring costs ($C_W/N=100$), the WS-type network ($\alpha=0$) is most vulnerable, and the robustness related to random failures undergoes an inversion between these two cost values, simply reflecting the non-trivial behavior of $\langle d \rangle (p)$. We observed an analogous effect in 2 dimensions.

In the case of overload, we found the behavior to be independent of the wiring costs of the initial network. The vulnerability always increases with $\alpha$, that is, the relative increase of the mean distance (caused by deleting a certain fraction of the most charged links) is lowest for $\alpha=0$. Tab.\ \ref{tab:ol} illustrates this result for $2D$ topologies with wiring costs of the initial networks $C_W/N=10$ and a removal fraction $q=10\%$. This finding is in agreement with the arguments given above in the context of flows. The higher $\alpha$ the smaller the fraction of real shortcuts carrying the majority of the traffic and making $\langle d \rangle$ small. As a consequence, $\langle d \rangle$ increases the faster upon their removal the larger $\alpha$. Furthermore, this result does not depend on the details of the overloading process: whether a given fraction of the most charged links is removed simultaneously or the failure is accomplished in a cascade-like fashion, the dependency of the robustness from $\alpha$ remains unaltered.

In summary, we have shown that small-world networks can be constructed in a very economical 
way if the parameters $D$ and $\alpha$ are chosen appropriately, also allowing for a more optimal distribution of 
costs over the nodes and flows over the links.
We also obtained the 
non-trivial picture that the WS-type SW network is most vulnerable if a large amount of wire is available.
In the case of overload, on the other hand, the length distribution alone fully determines the robustness, 
that is, networks characterized by a high value of $\alpha$ are most vulnerable. As length-distributions 
of the type investigated here have been observed in a number of real-world networks, such as integrated circuits, the Internet or the human cortex, 
we believe this work to have intriguing implications in their modelization.

We are grateful to Marc Barth\'elemy for his very valuable comments and discussions as well as to the EC-Fet Open project COSIN IST-2001-33555. This work has partially been supported by the Future and Emerging Technologies program of the EU under EU Contract 001907 DELIS. Both the COSIN and DELIS contracts have actually been supported by the OFES-Bern (CH).

\end{document}